\documentclass[preprint2]{aastex}
\usepackage{spr-astr-addons}
\usepackage{graphicx}
\begin{document}
\title{Two-flow magnetohydrodynamical jets around young stellar objects}

\shorttitle{Two-flow MHD jets around YSO}        

\author{Fabien Casse}
\affil{AstroParticule \& Cosmologie (APC)
10, rue Alice Domon et L\'eonie Duquet F-75205 Paris Cedex 13, France}
\email{fcasse@apc.univ-paris7.fr}
 
\author{Zakaria Meliani}
\affil{Institute for Plasma Physics ``Rijnhuizen''
P.O. Box 1207 NL-3430 BE Nieuwegein, Netherlands}
 
\author{Christophe Sauty}
\affil{Observatoire de Paris, L.U.Th F-92190 Meudon, France}

\begin{abstract}
We present the first-ever simulations of non-ideal
     magnetohydrodynamical (MHD)
     stellar winds coupled with disc-driven jets where the resistive and
     viscous accretion disc is self-consistently described. The
     transmagnetosonic, collimated MHD outflows  are investigated 
     numerically using the VAC code. Our simulations show that the inner
     outflow is accelerated from the 
     central object hot corona thanks to both the thermal pressure  and the
     Lorentz force. In our framework, the thermal acceleration is sustained 
     by the
     heating produced by  the  dissipated magnetic energy due to
     the turbulence. Conversely, the  outflow
     launched from the resistive accretion disc is mainly accelerated by
     the magneto-centrifugal force.  
     We also show that when a dense inner stellar wind
     occurs, the resulting disc-driven jet have a different structure,
     namely a magnetic structure where poloidal magnetic field lines are
     more inclined because of the pressure caused by the stellar wind. This
     modification leads to both an enhanced mass ejection rate in the
     disc-driven jet and a larger radial extension which is in
     better agreement with the observations besides being more consistent.
\end{abstract}

\section{Observational clues}
\label{intro}
The high velocity of the observed jet 
in YSOs suggests that they originate from a region that is no larger than
one astronomical unity (AU) in 
extent \citep{KwanTademaru88} and between 0.3 to  4.0 AU from the star in the
case of the LVC of DG Tau \citep{Andersonetal03}. 
  This theoretical prediction may be supported  for the disc wind
by the possible observations of the rotation of several jets associated
with TTauris  \citep{Coffeyetal04}
Moreover, in the case of Classical TTauris (CTTS)  
UV observations \citep{Beristainetal01,Dupreeetal05} reveal the presence of a 
warm wind which temperature  is at least of $3\times 10^5 {\rm K}$.
It also appears that the source of this wind is restricted to the star itself. 
These observations are supported
by X-ray observations  \citep{Feigelson&Montmerle99} that reveal the presence 
in CTTS of hot 
corona.  These observations also suggest the existence of stellar winds in CTTS
comparable to the solar wind. These winds may be thermally as well as magneto-centrifugally
accelerated.\\
The aim of the present work is to investigate the formation of two component 
outflows around YSOs, one coming from the thin accretion disc and the other
one being injected 
from the hot corona of the central star. This work is developed on the 
base of the  disc wind simulations of
 \citet{Casse&Keppens2002,Casse&Keppens2004} (hereafter CK04). 
The motivation is to study the influence of the stellar wind on the structure 
and the dynamics of the jet around YSOs.
\section{Two-flow jets around YSOs}
\label{sec:1}
\subsection{MHD simulations set-up}
In order to get the evolution of such a disc we solve, by mean of the VAC
code designed by \citet{Toth96}, the system of
time-dependent resistive and viscous MHD equations, namely, the usual
conservation of mass, momentum and total energy density $e$,
\begin{eqnarray}\label{Eq_dEnergydt}
e = \frac{\vec{B}^2}{2} + \frac{\rho \vec{v}^2}{2} + \frac{P}{\gamma - 1} &+&
\rho \Phi_{\rm G}  \\
\frac{\partial e}{\partial t} + \vec{\nabla} \cdot \left[ \vec{v} 
\left( e + P + \frac{B^2}{2}\right) - \vec{B} \vec{B} \cdot \vec{v}\right] 
&=&  \nonumber \\
\eta_m \vec{J}^2 - \vec{B} \cdot \left(\vec{\nabla} \times \eta_m \vec{J}\right)
&-& \nabla \left(\vec{v} \cdot\eta_{v} \hat{\Pi}\right) \nonumber
\end{eqnarray} 
where $\rho$ is the plasma density, $\vec{v}$ the velocity, $P$ the thermal
pressure, $\vec{B}$ magnetic field and $\vec{J} = \vec{\nabla}\times B$ is the
current density (provided through the MHD induction equation also solved by
the code). The gravitational potential is given by the classical
Newton potential generated by a central mass $M_*$. Note that both resistivity
($\eta_m$) and viscosity ($\eta_v$) are taken into account in the MHD set of
equations.  We adopt in our simulations a magnetic
Prandtl number $Pr = \frac{\eta_{v}}{\eta_{m}}=1$, believed to be an upper
limit of actual Prandtl number in YSO \citep{Lesu07}. Even with such high value,
we have 
demonstrated that the viscous torque is always much less efficient to remove
angular momentum than magnetic torque in thin disks
\citep{melianietal06}. The viscous-resistive disk structure is not modified
by the viscous torque as long as the Prandl number is $Pr\leq 1$. We thus 
introduce  similarly an anomalous viscosity $\eta_{v}$ equals to $\eta_{m}$. 
Through the dependence on the Alfv\'en velocity in our $\alpha$
prescription \citep{Shakura&Sunyaev73},
this becomes a  profile varying in time and space that essentially vanishes
outside the disc. We take $\alpha_{m}=0.1$ smaller than one to ensure that 
the Ohmic dissipation rate at the mid-plane of the accretion disc does not
exceed the rate of gravitational energy release \citep{Konigl95}. The
initial conditions as well as boundary conditions are fully displayed in 
\citet{melianietal06}.
\begin{figure*}[ht]
\begin{center}
\includegraphics[width=0.8\textwidth]{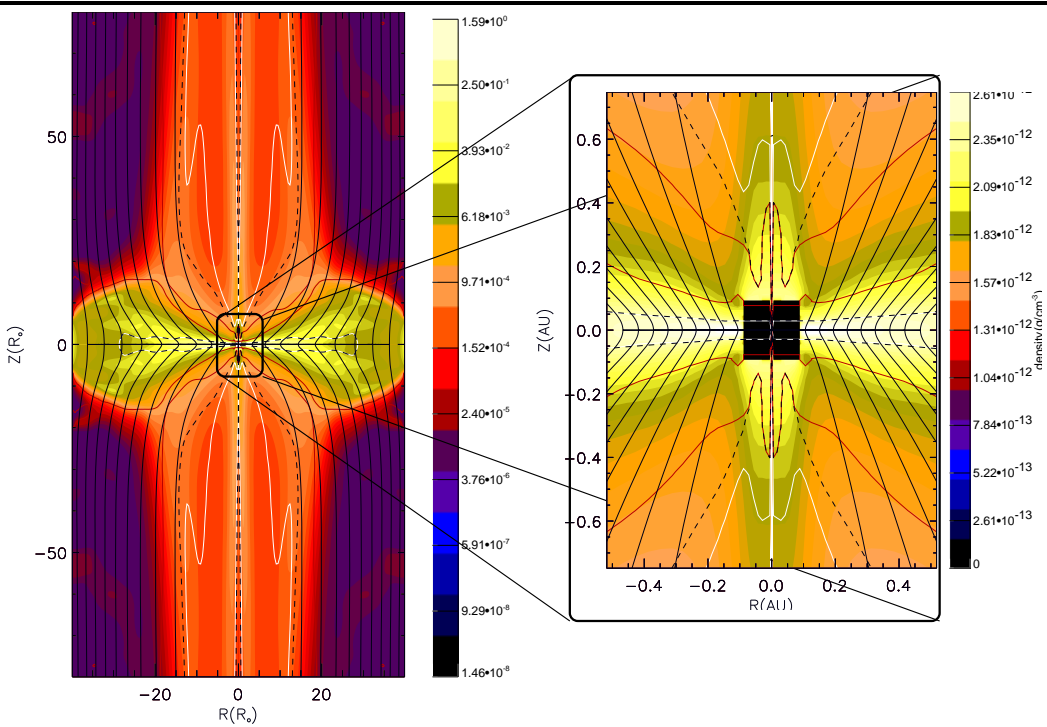}
\end{center}
\caption{Density contours in the poloidal plane of an
  accretion-ejection structure where a viscous and resistive MHD disc is
  launching a collimated jet. Magnetic field lines are drawn 
in black solid lines while the fast magnetosonic surface corresponds the
white solid line (Alfv\`en surface is the black dotted line and slow
magnetosonic surface correspond to the red line). A
  non-ideal stellar wind is emitted from the inner region which ejection mass
  rate is $\dot{M}=10^{-9}M_\odot/yr$.  The disc-driven jet conserves a
  dynamical structure very similar to the case where no stellar wind is
  emitted. }
\label{Fig1}
\end{figure*}
\subsection{Non-ideal MHD effects in stellar winds}
In most stellar wind models, the wind material is often subject to a
coronal 
heating, contributing to the global acceleration of the flow.
  In our simulations, we assume that the coronal heating is a 
fraction $\delta_{\varepsilon}$ of the energy released  in the 
accretion disc at the boundary of the sink 
region which is  transformed into thermal energy in the stellar corona close to the
 polar axis. 
This scenario was proposed by \citep{Matt&Pudritz05} and is
supported by the current observations of hot stellar outflows \citep{Dupreeetal05}. 
The $\delta_{\varepsilon}$ parameter range is limited, from below, by the initial thermal
acceleration  at the surface of the corona which should balance the
gravitational force and, from above, the condition to avoid a  too high
temperature in the corona (this gives the upper limit). In our simulation
we take a small efficient  
heating corona $\delta_{\varepsilon}=10^{-5}$. 
The interaction between the different components of the outflow is 
responsible for energy dissipation inside the plasma. This energy dissipation 
is the outcome of non-ideal MHD mechanisms occurring in the wind.  
In this paragraph, we show how these non-ideal MHD effects are taken into account to
 prescribe the magnetic resistivity
taking place in the wind region in addition to the disc resistivity
\begin{eqnarray}
\eta_{m} &=& \alpha_{m} \left.V_{A}\right\vert_{Z = 0} H \exp\left(-2
\frac{Z^2}{H^2}\right) \nonumber\\ &+& \alpha_{w} V_{A} H_{w}  \exp{\left[-2
\left(\frac{R}{H_{w}}\right)^2\right]} \ .
\end{eqnarray}
The first term accounts for the anomalous resistivity occurring in the
accretion disc. It vanishes outside the disc ($Z>H$).  The second
term corresponds to the description of an anomalous resistivity occurring in
the  outflow close to its polar axis. This term vanishes outside the stellar wind
($R>H_{w}$) where $H_{w}$ is the distance from the polar axis where the Alfv\`en
speed encounters  a minimum.
Hence, the dissipation effects are located in the stellar wind
component only and not in the disc wind which is supposed to be less turbulent.
For the resistivity in the stellar wind we take $ \alpha_{w}= 10^{-2}$, a
lower value than in the disc itself.
\subsection{Stellar wind embedded in a disc-driven jet}
\begin{figure*}[ht]
\begin{center}
  \includegraphics[width=\textwidth]{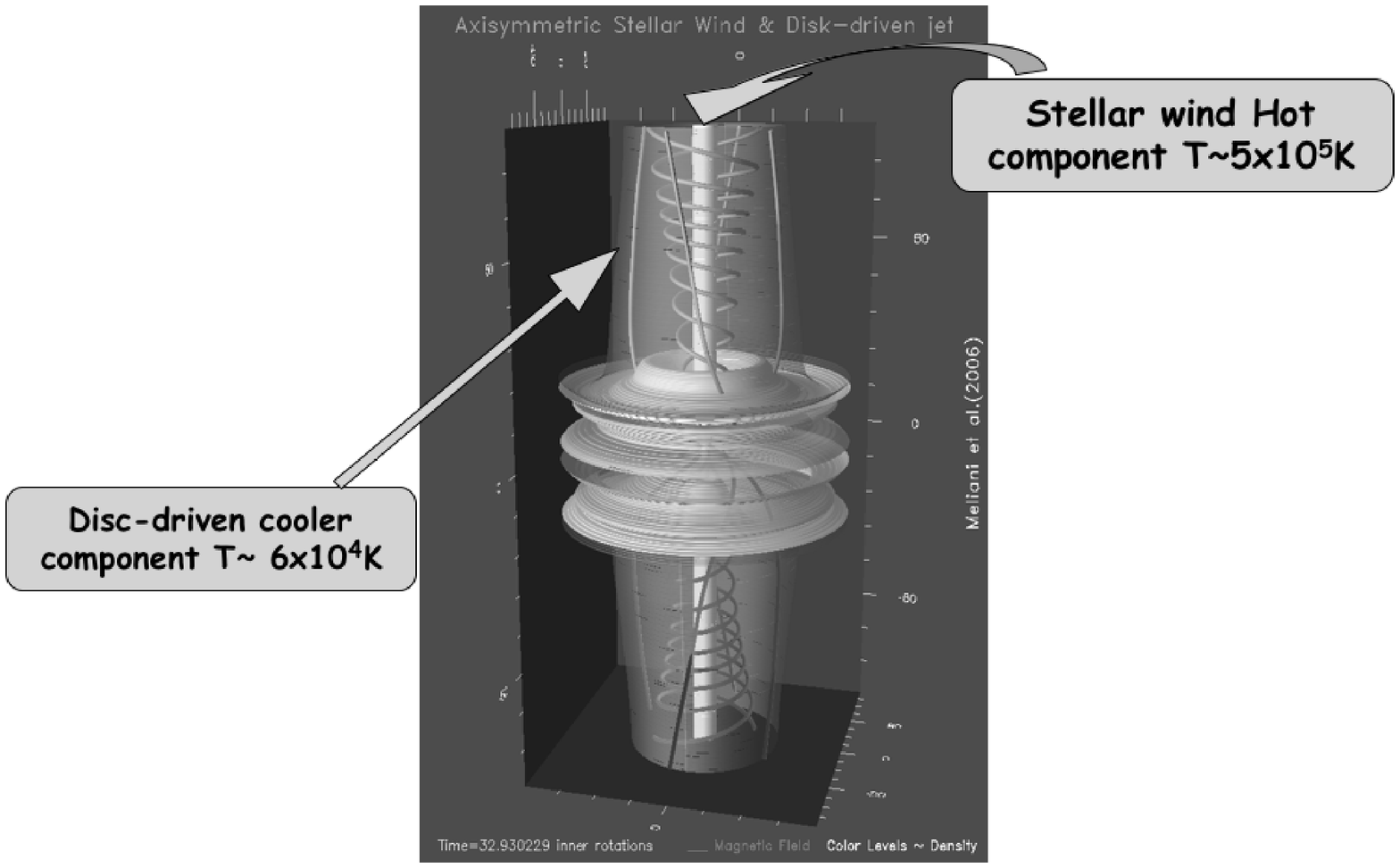}
\end{center}
\caption{Three-dimensional picture of a stellar wind embedded in a
  accretion disc-driven jet (cf Fig.~\ref{Fig1}). Colored surfaces stand for
  temperature 
  levels while green lines represent some magnetic field lines. We clearly see two
  components arising from this structure: a hot one related to the stellar
  wind while a more extended cooler one collimates the overall outflow.}
\label{Fig2}       
\end{figure*}
We first focus on a simulation where the stellar mass loss is set to
$10^{-9}M_{\odot}/yr$. The outcome of our simulation can be seen on
Fig.~\ref{Fig1} where we have displayed a snapshot of the poloidal
cross-section  of the structure.  In this snapshot we have
displayed respectively the density  contours (color levels) and the poloidal
magnetic field lines (solid lines).  The initial accretion disc
configuration is close to a hydrostatic equilibrium where the centrifugal
force and the total pressure gradient balance
the gravity. In the central region, the matter is continuously emitted at
the surface of the sink region (designed to be close to the star surface)
with sub fast-magnetosonic speed and with a solid 
rotation  velocity profile.  Initially, a conical hot outflow (stellar wind)
propagates
above the inner part of the disc. Its inertia compresses the magnetic field
anchored to the accretion disc. As a result the bending of the magnetic
surfaces 
increases, leading to a magnetic pinching of the disc. This pinching delays
the jet launching as the disc has to find a new vertical equilibrium. Thus the disc 
takes a few more inner disc rotations before launching its jet compared to CK04. 
Once the
jet has been launched the structure reaches a quasi steady-state where the
outflow becomes parallel to the poloidal magnetic field which is parallel
to the vertical direction.\\
The obtained solution is fully consistent with an accretion disc
launching  plasma  with a sub-slowmagneto\-sonic
velocity. The solution crosses the three critical surfaces, namely the slow-magneto\-sonic,
the Alfv\'en and the fast-magnetosonic surfaces. The
other component of the outflow, namely the stellar wind, is injected with
sub-fast\-magnetosonic velocity and crosses  the Alfv\'en and
fastmagnetosonic surfaces.  
The two components of the outflow become super-fastmagnetosonic  before
reaching the upper boundary limit of the computational
domain. Fig.~\ref{Fig1} also shows that
the outflow has achieved a quite good collimation within our computational
domain. We can  distinguish between the two components using the
isosurfaces of temperature which are displayed as color surfaces in
Fig.~\ref{Fig2}. In this figure, we can clearly see
a hot outflow coming from the central object embedded
in the cooler jet arising from the accretion disc.\\
\begin{figure}[t]
\begin{center}
{\rotatebox{0}{\resizebox{7.5cm}{10cm}{\includegraphics{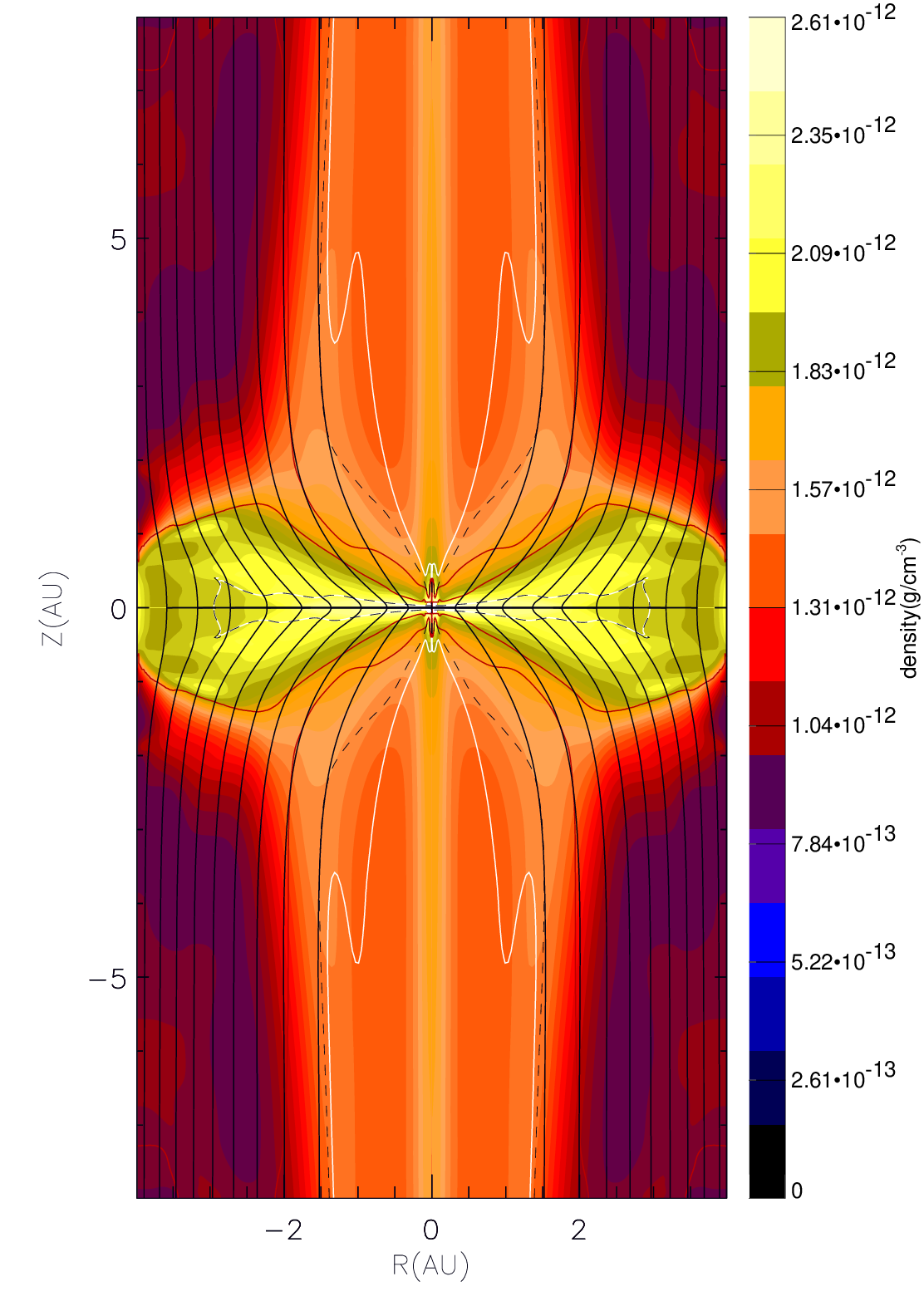}}}}
\end{center}
\caption{Same figure as in Fig.~\ref{Fig1} but with a
  non-ideal stellar wind emitted from the inner region which ejection mass
  rate $\dot{M}=10^{-7}M_\odot/yr$. The outflow structure is substantially
  modified by the 
  presence of the stellar outflow since its radial extension is two times
  larger than
  in the case with no or weak  stellar outflow. { The size of the sink region
is $R_i=0.1 {\rm AU}$ and the stellar mass is $1 M_{\odot}$.}}
\label{Fig3}
\end{figure}
In order to study the time evolution of both accretion and ejection
phenomena in the  accretion disc and around the star, we
analyzed  the  accretion and ejection mass loss rate in both components.
Similarly to CK04 we observe a strong increase of
the accretion rate in the inner part with time. This behaviour is related to the
extraction of  the rotational energy of the accretion disc by the magnetic
field. Indeed the creation of the toroidal component of the magnetic field
in the disc brakes the disc matter so that the centrifugal force decreases
 leading to an enhanced accretion motion. The mass flux associated with the
 disc-driven jet slowly increases to reach $18\%$
of the accreted mass rate at the inner radius and contributes to $98\% $ of the
total mass loss rate of the outflow. In fact, in this simulation the mass loss
rate from the central object is constant ($10^{-9}M_{\odot}/yr$) while the
inner accretion rate reaches 
$10^{-6}M_{\odot}/yr$ and the disc-driven jet mass rate
$10^{-7}M_{\odot}/yr$. Hence the stellar outflow does not affect much the
overall structure of the outflow. This is confirmed by the shape of the
outflow since it is reaching a very similar aspect to the one obtain  in CK04 or
in the previous simulation without a stellar jet, i.e. a jet
confined within $20$ inner disc radius.\\
 On Fig.~\ref{Fig1}, we have displayed density isocontours within
  a small area around the sink region. Thanks to this plot, we can see that
  the magnetic lever arm associated with the various outflow components are
  different. Indeed the disc-driven jet exhibits magnetic lever arm
  (related to the ratio of the Alfv\`en radius to the magnetic field line
  foot-point radius) 
  varying approximately from $9$ to $25$ while the magnetic lever arm associated with the
  stellar wind is ranging from $0$ near the axis to several tens, if one
  considered the foot-point of magnetospheric magnetic field line to be
  anchored to the star. This last magnetic lever arm value may not be very
  reliable since we have imposed the size of the sink and thus influenced
  the radial extension of the magnetospheric outflow near the
  sink. Nevertheless, this simulation illustrates the fact that such
  large stellar wind magnetic lever arm may be compatible with two-component
  collimated outflow launched from YSO. This may be the physical mecanism
  responsible for
  stellar braking that has to occur in many low-mass stars (see
  e.g. \citet{Matt&Pudritz05}). One important issue remains in this
  model: the amount of thermal energy released by ohmic heating in the
  stellar wind is of the order of $35\%$ of the energy released by
  accretion. One has then to explain how such an amount of energy may be
  carried away by MHD turbulence to heat the stellar wind. This question
  remains open.  
\subsection{Massive stellar winds vs. sun-like mass loss rate wind 
effects on disc-driven jet}
In the simulations presented so far, we have seen  that  winds
with mass loss rate similar to the Sun (up to $10^{-9}M_\odot/yr$)
do not greatly influence the disc outflow since their
general behaviour remains similar. However in the case of a massive
stellar jet,  the inner wind may strongly influence the outflow as it can be 
seen in a 
new simulation performed for a stellar wind mass loss rate  set to
$10^{-7}M_{\odot}/yr$ (Fig.~\ref{Fig3}).
The radial stellar wind compresses strongly the magnetic field anchored in the 
accretion disc. The enhanced magnetic field bending (even in the external part 
of the accretion disc $R>30$) leads to an increase of the magnetic pinching in 
an extended region of the disc $1<R<30$. Thus the outflow is launched from all 
this region since the Blandford \& Payne criterion is fulfilled everywhere
\citep{BP82}.  Indeed the magnetic field becomes dynamically dominant in the 
disc corona of this region. The magnetic bending larger than $30^{\circ}$
from the vertical direction leads to a centrifugal force and a thermal
gradient pressure more efficient to launch the outflow from the disc 
as it can be seen in the jet mass loss
which reached $0.5$ of the accretion rate in the inner part.\\
The angular momentum carried away by the stellar outflow now  represents $5\%$ of 
the accreted angular momentum at the inner radius of the accretion disc. 
Regarding the acceleration of the outflow, we can distinguish two regions:
an internal one corresponding to the contribution from the stellar outflow
and an external one coming from the disc-driven
jet. 
\section{Outlook}
The present work tried to illustrate the ability of accretion disc-driven
jets to focus outflows coming from the central object. We applied MHD
simulations to demonstrate this statement in the context of YSOs by
self-consistently describing both the accretion disc, its related outflow
{\it and} the wind acceleration. These simulations however left unanswered
important question regarding the stellar coronal heating that has to occur
in order to give birth to the stellar wind. Another question is the origin of the
turbulence warming up the stellar wind as this is also a problem in solar physics.\\
 It is noteworthy that this configuration may be useful in other contexts as for
instance in microquasars and AGN. Indeed, when a black hole is the central
object of the system, relativistic outflows are believed to arise from its
ergosphere (see, e.g. the contribution of J. McKinney in these
proceedings). This kind of flow is prone to a substancial decollimating
force originating from the displacement current occurring in relativistic
MHD regime \citep{Bogovalov&Tsinganos05}. The collimating action provided
by a large-scale non-relativistic disc-driven jet would then be useful to
explain the collimation of jets observed in AGN and microquasars.


\bibliographystyle{natbib}

\bibliography{sample}


\end{document}